\documentclass[modern]{aastex63} 
\usepackage{amsmath}

\lefthead{Wada, Tsukamoto, Kokubo}
\righthead{Planets around SMBH}

\newcommand{\bea}{\begin{eqnarray} }
\newcommand{\eea}{\end{eqnarray}}

\begin{document}


\title{Formation of ``Blanets" from Dust Grains around the Supermassive Black Holes in Galaxies}
 
\author{%
Keiichi Wada
}
\affiliation{Kagoshima University, Graduate School of Science and Engineering, Kagoshima 890-0065, Japan}
\affiliation{Ehime University, Research Center for Space and Cosmic Evolution, Matsuyama 790-8577, Japan}
\affiliation{Hokkaido University, Faculty of Science, Sapporo 060-0810, Japan}
\correspondingauthor{Keiichi Wada}
\email{wada@astrophysics.jp}

\author{
Yusuke Tsukamoto
}%
\affiliation{Kagoshima University, Graduate School of Science and Engineering, Kagoshima 890-0065, Japan}

\author{Eiichiro Kokubo}%
\affiliation{National Astronomical Observatory of Japan, Mitaka 181-8588, Japan}


%


\begin{abstract}
In Wada, Tsukamoto, and Kokubo (2019), we proposed for the first time that a new class of planets, \textit{blanets}, 
can be formed around supermassive black holes (SMBHs) in the galactic center. 
Here, we investigate  the dust coagulation processes and physical conditions of 
the blanet formation outside the snowline ($r_{snow} \sim $ several parsecs) in more detail, especially considering 
the effect of the {radial drift} of the dust aggregates. 
{We found that a dimensionless parameter $\alpha = v_t^2/c_s^2$, where $v_t$ is the turbulent velocity 
and $c_s$ is the sound velocity, describing  the turbulent viscosity should be smaller than 0.04 in the circumnuclear disk } 
to prevent the destruction of the aggregates due to collision.
The formation timescale of blanets $\tau_{GI}$ at $r_{snow}$ is, $\tau_{GI} \simeq$ 70-80 Myr for $\alpha = 0.01-0.04$ and $M_{BH} = 10^6 M_\odot$.
The mass of the blanets ranges from $\sim 20 M_E$  to $3000 M_E$ in $r < 4$ pc for $\alpha = 0.02$ ($M_E$ is the Earth mass),
 which is in contrast with $ 4 M_E-6 M_E$ for the case without the {radial drift}. 
Our results suggest that {blanets} could be formed around relatively low-luminosity AGNs ($L_{bol} \sim 10^{42}$ erg s$^{-1}$)
 during their lifetime ($\lesssim 10^8$ yr). 
\end{abstract}


\section{INTRODUCTION}
There is enough evidence suggesting that 
planets are formed in the {circumstellar} disks around stars. 
However,  stars might not be the only site for planet formation.
Recently, in Wada, Tsukamoto, and Kokubo (2019) (hereafter Paper I), we claimed 
a new class of {``planets"} \footnote{{Here, we merely call massive rocky/icy objects orbiting around a central gravity source as ``planets''. }}
that orbit around super-massive black holes (SMBHs) in 
galactic centers.
Paper I theoretically investigated the growth processes of planets, from sub-micron-sized icy dust
 monomers to Earth-sized bodies outside  the snowline in a circumnuclear disk around a SMBH, typically located several 
 parsecs from the SMBHs.
As is the case in a protostellar disk,
 in the early phase of the dust evolution, low-velocity collisions between dust particles promote
 sticking; therefore, the internal density of the dust aggregates
 decreases with growth \citep{okuzumi2012, kataoka2013}.
When the size of porous dust aggregates reaches 0.1--1 cm, the collisional and the gas-drag compression become effective, and as a result, the
 internal density stops decreasing.
Once 10--100 m sized aggregates are formed, they decouple from gas 
 turbulence, and as a result, the aggregate layer becomes gravitationally unstable \citep{michikoshi2016, michikoshi2017},
 leading to the formation of ``planets" due to the fragmentation of the layer, with ten times the mass of the earth.
The objects orbit the SMBHs with an orbital time of $10^5-10^6$ years.
 To distinguish them from standard planets,  we hereafter call these hypothetical astronomical objects  \textit{blanets} \footnote{{This does not necessarily mean a simple abbreviation of 
``black hole planet'', because this new class of objects does not resemble the planets in the solar system nor any known exoplanet systems, in a sense that a swarm of super-Earth mass objects are orbiting around the central gravity source.  See also \S 4.2.}}.

 The results reported in Paper I, however, have two major limitations. 
 One is that the collisional velocity between the dust aggregates might become too large ($>$ several 100 m s$^{-1}$
at the Stokes parameter, $S_t \sim 1$).
 And if the collisional velocity is that large, rather than growing, the aggregates might get destroyed.
 In Paper I, we used the numerical experiments conducted by \citet{wada-k2009}\footnote{Note that \citet{wada-k2009} and 
 \citet{wada-k2013} were written by Koji Wada and his collaborators, 
 not by the first author of this paper.} on the collisions between the dust aggregates, wherein 
 the critical collisional velocity ($v_{crit}$) scales with the mass $m_d$ of the dust aggregates, as
 $v_{crit} \propto m_d^{1/4}$. However, this is correct only for the head-on collisions, as stated in the paper.  
Moreover, \citet{wada-k2009, wada-k2013} showed
that the growth efficiency of the dust aggregates depends on the impact parameter of the collisions, 
and as a result,  $v_{crit}$ does not strongly depend on the mass of the dust aggregates, if off-set collisions are taken into account. They concluded that
 $v_{crit} \simeq 80$ m s$^{-1}$ for the ice monomers\footnote{{Here, we suppose water ice \citep{sato2016}. 
 In fact, the presence of H$_2$O in AGNs is suggested by maser observations \citep[e.g.,][]{greenhill2003}, and by chemical models \citep[e.g.,][]{wada2016}. Note that $v_{crit}$ is much smaller ($\sim 1$ m s$^{-1}$) for silicate monomers \citep{wada-k2009}). Therefore, we here consider dust evolution outside the snowline.  }}.
 This low critical velocity is also one of the obstacles in the planet formation in circumstellar disks.
In this follow-up paper, we adopt $v_{crit} \simeq 80$ m s$^{-1}$ as a constraint on the growth of the dust aggregates.
 
Another limitation of Paper I is that the size of dust aggregates $a_d$ and collisional velocity $\Delta v$ show 
runaway growth in the collisional compression phase around $S_t \sim 1$. 
However, this rapid growth would not be realistic if a more natural treatment of the internal density of the dust is considered (\S2.2.1, see also \S 3).

Moreover, there is a critical process that may promote blanet formation.
In paper I, we did not take into account the radial drift of the dust particles as the first approximation.
The radial velocity of the dust $v_{r, d}$ relative to the gas \citep{weidenschilling1977, tsukamoto2017} is
$v_{r, d} \sim S_t \, \eta \, v_K$, and $\eta \sim (c_s/v_K)^2$, where $c_s$ is the gas {isothermal} sound velocity and $v_K$ is 
the Keplerian rotational velocity. 
In the circumnuclear disk around a SMBH, initially $S_t \, \eta \sim 10^{-4} - 10^{-3}$. 
Then the drift time of the dust particle $t_{drift} \sim r/v_{r,d} \sim 5-50 (M_{BH}/10^7 M_\odot) ^{1/2} (r/ 1 \, {\rm pc})^{1/2}$ Myr.
This is not negligibly small for the lifetime of the active galactic nucleus (AGN), i.e., $10^7- 10^8$ yr.
In this paper, we investigate the effects of the {radial drift} of
the dust particles.

 The remainder of this paper is organized as follows. In \S 2, we describe the models for the dust evolution and 
 its application to the circumnuclear region. 
  In \S 3, we show the results of the models with and without the {radial drift} of the dust particles.
In \S 4, we discuss how the maximum collisional velocity and the formation timescale of blanets 
depend on the parameters $\alpha$ and $M_{BH}$.  
We also discuss the expected mass of the blanets and their radial distribution.
Finally, we summarize the results in \S 5. 


\begin{figure}[h]
\begin{center}
\includegraphics[width = 10cm]{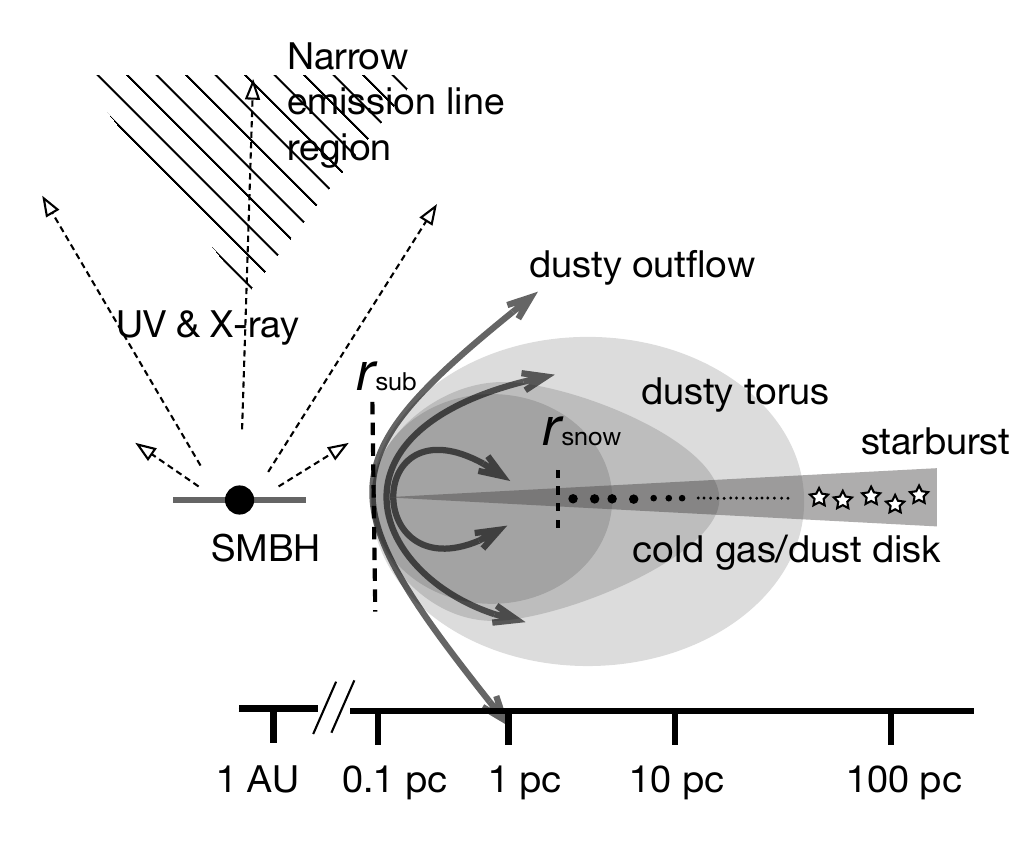}    
\caption{
A schematic picture of the Active Galactic Nucleus (AGN) and the circumnuclear disk.}
\end{center}
\label{fig:1}
\end{figure}

%
\section{Models}
\subsection{The region of ``blanet formation''}

Here we briefly summarize the concept of dust evolution around SMBHs, as discussed in Paper I. 
Figure \ref{fig:1} shows a schematic of the active galactic  nucleus (AGN) and the circumnuclear disk. A SMBH (with a mass of $10^6-10^{10} M_\odot$) is surrounded by an accretion disk, which radiates enormous energy (the bolometric luminosity is $\sim 10^{42}- 10^{45}$ erg s$^{-1}$), mostly as  ultra-violet and X-rays.  The dust particles in the central $r < r_{sub} $ are sublimated by {the radiation from the accretion disk around the SMBH}.
The radius  depends on the AGN luminosity:

\begin{eqnarray}
r_{sub} \simeq 1.3 \, {\rm pc} \left( \frac{L_{UV}}{10^{46} \, {\rm erg}\, s^{-1}}\,  \right)^{0.5} \; \left(\frac{T_{sub}}{1500 \, K} \right)^{-2.8}  
 \; \left(\frac{a_{d}}{0.05 \, \mu m} \right)^{-1/2}.
\end{eqnarray}

where $L_{UV}$ is the ultra-violet luminosity of the AGN,  and $a_d$ is the dust size  \citep{barvainis1987}.
The radiation forms conical ionized gas (narrow emission-line region) and also contributes to producing the outflows of the dusty gas and torus \citep{wada2012, wada2018, izumi2018}.
In the mid-plane of the torus, cold, dense gas forms a thin disk, where icy dust particles can be present beyond the snowline $r_{snow}$ (see \S 2.4). 

{
We introduce a free parameter $\alpha \equiv v_t^2/c_s^2$, 
where $c_s$ is the gas sound velocity and $v_t$ is the turbulent velocity,  to represent strength of the kinematic viscosity due to the turbulence
\citep[e.g.,][see also eqns. (\ref{eq: deltav1}) and (\ref{eq: deltav2})]{ormel2007}.
}
In the circumnuclear disk in AGNs, the value of $\alpha$ is highly uncertain\footnote{In AGNs, the turbulence could be generated by various mechanisms; e.g., the magneto-rotational instability \citep{kudoh2020}, 
the self-gravity \citep{shlosman1987}, the radiation-driven fountain \citep{wada2012}, and the stellar feedback \citep{wada2002b}.}. 
Therefore, we here treat $\alpha$ as a free parameter to check how it alters the results, especially 
the maximum collisional velocity between the dust aggregates and 
the onset of the gravitational instability of the dust disk.

In contrast to the dust coagulation process in the circumstellar disks \citep{weidenschilling1977}, 
the drag between dust particles and gas obeys the Epstein law.  
The aggregate's size ($a_{d}$) is always much smaller than the mean free path of the gas, 
$\lambda_g \sim   10^{12} \; {\rm cm} ( {\sigma_{mol}}/{10^{-15} \, {\rm cm}^2} )^{-1}
( {n_{mol}}/{ 10^3 \, {\rm cm}^{-3}} )^{-1}$, where $\sigma_{mol}$ and $n_{mol}$ are the collisional cross-section and
number density of the gas, respectively.

\subsection{Evolution of dust aggregates in each stage}

The model for the growth of dust particles here is based on
 the elementary processes found around stars.
The evolution of dust particles is divided into four stages as described below.

\subsubsection{Hit-and-Stick stage}
If the dust aggregates grow through ballistic cluster-cluster aggregation (BCCA), 
the internal structure of the aggregate should be porous (i.e., the internal density is much smaller than the monomer's density: $\rho_{int} \ll \rho_0$), and 
its fractal dimension is $D \simeq 1.9$ \citep{mukai1992, okuzumi2009}.
This is called the \textit{hit-and-stick} stage \citep{okuzumi2012, kataoka2013}, and the internal density is given by
\begin{eqnarray}
\rho_{int} =  \rho_0 \left( \frac{m_d}{m_0} \right)^ {1-3/D},
\end{eqnarray}
where  $m_d$ is the mass of the aggregate, and $m_0$ is the monomer's mass.
We assume that $m_0 =10^{-15}$ g and $\rho_0 = 1$ g cm$^{-3}$.

The growth rate for $m_d$ is 
\begin{eqnarray}
\frac{d m_d }{d t} = \frac{2 \sqrt{2 \pi} \, \Sigma_d  \, a_d^2 \, \Delta v}{H_d}, 
\end{eqnarray}
where $\Delta v$ is collision velocity between the aggregates, and $H_d$ is the scale height of the dust disk as given in 
\citep{youdin2007, tsukamoto2017};

\begin{eqnarray}
H_d = \left( 1 + \frac{S_t}{\alpha} \frac{1+ 2 S_t}{1+ S_t} \right)^{-1/2}  H_g,
\label{eq: 5}
\end{eqnarray}
where $H_g = c_s/\Omega_K$ is the scale height of the gas disk, and 
{
$S_t$ is the Stoke parameter, i.e., the normalized stopping time is defined as $S_t \equiv t_{stop}/t_L$ with the eddy turn over time $t_L$.
Here we suppose that $t_L = \Omega_K^{-1}$ and
\begin{eqnarray}
S_t = \frac{\pi \rho_{int} \, a_d }{2 \Sigma_g}.
\label{eq:stokes}
\end{eqnarray}
}

The collision velocity between aggregates $\Delta v$ for $S_t < 1$ can be divided into  two regimes \citep{ormel2007}:
regime I) $ t_s \ll t_{\eta} = t_L \, Re^{-1/2}$, and  regime II) $t_\eta \ll t_s \ll \Omega_K^{-1}$.
Here the Reynolds number, 
$R_e \equiv \alpha c_s^2/(\nu_{mol} \Omega_K)$ with the molecular viscosity $\nu_{mol} \sim \frac{1}{2} c_s \lambda_g$ is
\begin{eqnarray}
R_e &\approx& 3 \times 10^4 \, \left(\frac{M_{BH}}{10^6 \, M_\odot} \right)^{-1/2} \, \left(\frac{r}{1 \, {\rm pc}} \right)^{3/2} \, c_{s, 1}^{-1} \, Q_g\,  \left(\frac{\gamma_{Edd}}{0.01}\right),
\end{eqnarray}
where $Q_g$ is the Toomre's $Q$-value for the gas disk and $\gamma_{Edd}$ is the Eddington ratio for the AGN.
For the hit-and-stick stage,  $S_t  \ll  R_e^{-1/2}$, then for the regime I, 
\begin{eqnarray}
\Delta v_I \simeq \sqrt{\alpha} c_s R_e^{1/4} |S_{t,1} - S_{t,2}|
=
C_I \sqrt{\alpha} c_s R_e^{1/4} S_{t}, 
\label{eq: deltav1}
\end{eqnarray}
where $S_{t,1}$ and $S_{t,2}$ are Stokes numbers of the two colliding particles, and 
$C_I$ is a constant of the order of unity \citep{sato2016}.
For regime II, on the other hand, 
\begin{eqnarray}
\Delta v_{II} &\simeq & v_L \sqrt{t_{stop}/t_L} \simeq \sqrt{\alpha S_t} \, c_s  
\label{eq: deltav2}
\end{eqnarray}
where $v_L$ is velocity of the largest eddy.  
{We assume that $\Delta v_{I} = \Delta v_{II}$ at the transition. }

The size of dust aggregates determines how they interact with the gas.
The dynamics of the aggregates is affected by their cross sections, which depend on
their internal inhomogeneous structure.
The radius of BCCA cluster $a_{BCCA}$ consisted of $N$ monomers ($N = m_d/m_0$) is given as 
 $a_{BCCA} \simeq N^{0.5} a_0$ for $N \gg 1$ \citep{mukai1992, wada-k2008a, wada-k2009}, which was also confirmed through $N$-body simulations \citep{suyama2012}. We then assume that 


\begin{equation}
a_d (m_d) = \begin{cases}
     \left(\frac{m_d}{m_0} \right)^{1/2}  a_0  & (BCCA) \\
     \left(\frac{3 m_d}{4 \pi \rho_{int}} \right) ^{1/ 3} & (otherwise).
  \end{cases}
  \label{eq:10}
\end{equation}


\subsubsection{Compression stages}

The hit-and-stick stage ends due to collisions between the aggregates (collisional compression), or due to their
interaction with the ambient gas (gas-drag compression).
In the collisional compression,  the rolling energy $E_{roll}$, which is the energy required to rotate a particle around a connecting point by 90$^\circ$,
is comparable to the impact energy, $E_{imp} = m_d \Delta v^2/4$, between the two porous dust aggregates of the same mass, $m_d$.  Beyond this point,  the aggregates start to get compressed
due to mutual collisions and interaction with the gas (i.e., the ram pressure).

According to  \citet{suyama2012}, the internal density of the aggregated $\rho_{int, f}$ formed by collisions between two equal-mass aggregates, 
with density $\rho_{int}$, is calculated for $E_{imp} > 0.45 E_{roll}$ : 
\begin{eqnarray}
\rho_{int, f} = \left( \rho_{int}^4  +  \rho_f^4 \, \frac{E_{imp} -0.45 E_{roll}}{0.15 N E_{roll}} \right)^{1/4},
\end{eqnarray} $\rho_f$ is the fractal density of the dust aggregate: $\rho_f \approx m_d/(7.7 a_d^{2.5})$, and
$E_{roll} = 4.37\times 10^{-9}$ erg. 

Moreover, the fluffy dust aggregates can be compressed owing to the ram pressure of the ambient gas \citep{kataoka2013}.
The internal density of the aggregates that are compressed by the gas is given
\begin{eqnarray}
\rho_{int, drag} \simeq \left(\frac{a_0^3}{E_{roll}} P_g \right)^{1/3} \rho_0,
\end{eqnarray}
where the ram pressure for a dust aggregate is 
\begin{eqnarray}
P_{g} =  \frac{m_d \, v_d}{\pi  a_d^2 \, t_s},
\end{eqnarray}
with the stopping time $t_s = S_t/\Omega_K$  \citep{kataoka2013}. 


As the aggregates become more massive ($m_d > 10^{10}$ g),
they start getting compressed owing to their self-gravity, and 
the internal density evolves as 
$\rho_{int} \propto (\Delta v)^{3/5}  \, m_d^{-1/5}$ \citep{okuzumi2012}. 


 \subsubsection{$N$-body stage}
 \label{subsec: nbody}
 {When $S_t \simeq 1$,
kinematics of the aggregates is affected not only by the turbulence, 
but also by mutual interaction between the aggregates as a N-body system
and  by the gravitational interaction with the density fluctuation due to the turbulence.
Then the collision velocity between the aggregates is determined by 
a balance between various heating and cooling processes as the $N$-body particles.}
According to \citet{michikoshi2016, michikoshi2017},   we solve the following equation to get the equilibrium random velocity of 
the dust aggregates $v_d$, 
\begin{eqnarray}
\frac{d v_d^2}{dt} &=& \left( \frac{d v_d^2}{dt} \right)_{grav} + \left( \frac{d v_d^2}{dt} \right)_{turb, stir} +  \left( \frac{d v_d^2}{dt} \right)_{turb, grav} \nonumber \\ 
&-& \left( \frac{d v_d^2}{dt} \right)_{coll}   -\left( \frac{d v_d^2}{dt} \right)_{drag} = 0 . 
\label{eq:19}
\end{eqnarray}
The first three heating terms represent the gravitational scattering of the aggregates, stirring by the turbulence, and gravitational scattering
by density fluctuation of the turbulence, respectively.  The two cooling terms in eq.(\ref{eq:19})  represent the collisional damping and the gas drag.  
{We assume the collision velocity $\Delta v \approx v_d$ at $S_t = 1$ and numerically solve eq. (\ref{eq:19}) in this stage.}

 \subsubsection{Radial drift of the dust particles}

In Paper I, we ignored the {radial drift} of the dust particles in the disk.  However, as mentioned in \S 1, this is not always obvious. 
Here, we solve the following governing equations for the dust evolution based on the assumption that the mass 
distribution of the dust particles at each orbit radius is singly peaked at a mass  \citep{tsukamoto2017, sato2016};

\begin{eqnarray}
\frac{\partial \Sigma_d}{\partial t} + \frac{1}{r} \frac{\partial}{\partial r} (r v\, _{r, d} \, \Sigma_d) &=&  0,  \\
\frac{\partial m_d}{\partial t}  + v_{r,d} \frac{\partial m_d}{\partial r}  &= & \frac{m_d}{t_{coll}} \, .     
\label{eq:advec}
\end{eqnarray}
Here $t_{coll}$ in eq.(\ref{eq:advec}) is the collision time, and the source term is 
\begin{eqnarray}
\frac{m_d}{t_{coll}} = m_d (4 \pi a_d^2 \, n_d \, \Delta v) = 2\sqrt{2 \pi} a_d^2 \, \Sigma_d \, \Delta v  H_d^{-1},
\end{eqnarray}
 where $n_d$ is the number density of the dust particles.

The dust particles have a radial velocity due to the drag with the ambient gas:

\begin{eqnarray}
v_{r,d} = -\left( \frac{v_{r,g}}{ 1 + S_t^ 2} + \frac{2 S_t}{ 1 + S_t^2}  \, \eta  \, v_K \right),
\end{eqnarray}
where $v_K$ is the Kepler velocity and $\eta$ is a parameter that determines the 
sub-Keplerian motion of the gas, and the radial velocity of the gas $v_{r, g}$ is given with
the mass accretion rate $\dot{M}$:
\begin{eqnarray}
v_{r, g} = - \frac{\dot{M} }{ 2\pi  r  \Sigma_g},
\end{eqnarray}
where the mass accretion rate $\dot{M}$  is assumed to be using 
the Eddington mass accretion rate
$ \dot{M} = \gamma_{Edd} \, \dot{M}_{Edd}$ with the Eddington ratio $\gamma_{Edd}$. 
 
 \subsection{Gravitational instability of the dust disk and formation of blanets}
We investigate the gravitational instability (GI) of the disk consisting of  
dust aggregates with $S_t > 1$ using
 the Toomre's $Q$-value defined as  
 \begin{eqnarray}
 Q_d \equiv \frac{(v_d/\sqrt{3}) \Omega_K}{3.36 G \Sigma_d}. 
\end{eqnarray}

For the axi-symmetric mode, $Q_d < 1$ is the necessary condition for GI,  but the non-axisymmetric mode can develop
 for $Q_d \lesssim 2$.
In this case, spiral-like density enhancements are formed followed by fragmentation of the spirals \citep{michikoshi2017},
which leads formation of massive objects, i.e., \textit{blanets}.
  The mass of blanets is estimated as 
 \begin{eqnarray}
 M_{bl} \simeq \lambda_{GI}^2 \Sigma_d, 
 \end{eqnarray}
where the critical wavelength for GI is
\begin{eqnarray}
\lambda_{GI} = \frac{4 \pi^2 G \, \Sigma_d}{\Omega_K^2}.
\label{eq: lambda_gi}
\end{eqnarray}


\subsection{Initial and boundary conditions}

In all the models,  the circumnuclear cold gas disk embedded in the geometrically thick torus (see Fig. \ref{fig:1}) is assumed to be gravitationally stable;
the Toomre's Q-value, $Q_g  \equiv c_s \Omega_K/\pi G \Sigma_g = 2$.  The gas sound velocity is assumed  to be  $c_s^2 = k_B T_g/\mu \, m_H $ with $T_g = 100$ K and $\mu = 2.0$.

The Eddington ratio is assumed to be $\gamma_{Edd} = 0.01$.   The AGN bolometric luminosity is 
then $L_{bol} = 1.3 \times 10^{42}  {\rm erg}\,  {\rm s}^{-1} M_{BH} / (10^6  M_\odot)$.
The X-ray luminosity of the AGN, which is used to determine the snowline radius, $L_X = 0.1 L_{bol}$. 
This can be attributed to the fact that the UV flux from the accretion disk is attenuated in the dense circumnuclear disk. 

The snowline for $a_d = 0.1 \, \mu m$, 
  \begin{eqnarray}
  r_{snow} \approx  1.5  \, {\rm pc}  \left(\frac{L_X}{1.3\times 10^{41} \, {\rm erg \,s}^{-1} }  \right) ^{1/2}  \left( \frac{T_{ice}}{170 \, {\rm K}} \right)^{-2.8} \, \left( \frac{a_{d}}{0.1\, \mu {\rm m}}  \right)^{-1/2}.  
  \end{eqnarray}  
 Therefore, it  is expected that the dust in the most part of the circumnuclear disk is icy.  We assume $T_{ice} = 170$  K.
 The dust to gas mass ratio is assumed as $f_{dg} = 0.01$.
   We solve the governing equations (\S 2.2.4) between $r =0.1$ pc and  200 pc with 600 grid cells.

\section{Resuts}

\begin{figure}[ht]
\begin{center}
\includegraphics[width = 16cm]{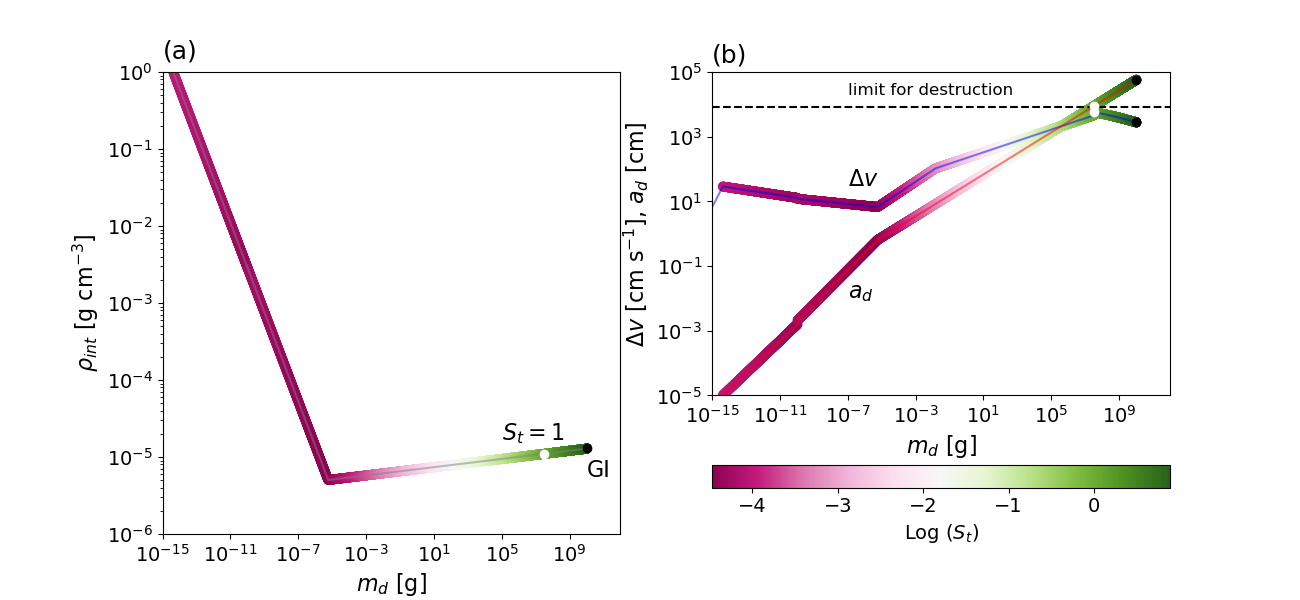} 
\caption{(a) Evolution of the internal density of a dust aggregate $\rho_{int}$ at the snowline ($r =$ 1.5 pc) 
as a function of the aggregate mass $m_d$ for $M_{BH} = 10^6 M_\odot$ and $\alpha = 0.02$ . 
Evolution prior to the gravitational instability (i.e., $Q_d > 2$) is plotted.
The positions where $S_t$ becomes unity and $Q_d = 2$  are shown by the filled white and black circles, respectively.
The color bar represents the Stokes number. 
\\
(b) Same as (a), but for collision velocity of the aggregates $\Delta v$ and size of the aggregate $a_d$. The dashed line shows
$\Delta v =$ 80 m s$^{-1}$, which is the limit for the collisional destruction of the aggregates suggested by numerical experiments \citep{wada-k2009}.
After $S_t = 1$ is attained, $\Delta v$ drops and the disk of the aggregates becomes gravitationally unstable. 
} \label{fig:2}
\end{center}

\end{figure}

Figure \ref{fig:2}a shows a typical evolution of  a dust aggregate  at the snowline
for $M_{BH} = 10^6 M_\odot$ and $\alpha = 0.02$.  
The internal density of the aggregate $\rho_{int}$ is 
plotted as a function of its mass $m_d$.
Initially, the internal density decreases monotonically from the monomer's initial density, i.e., $\rho_0 = 1$ g cm$^{-3}$
to $4 \times 10^{-6}$ g cm$^{-3}$, as its mass increases from $m_d \sim 10^{-15}$ g to $ \sim 10^{-5}$ g. 
At that instant, the size of the aggregate becomes $\sim 1$ cm (see Fig. \ref{fig:2}b). 
 After this hit-and-stick phase, the fluffy dust aggregates keep growing due to collisions
in the turbulent gas motion until  $S_t \simeq 1$.
During this stage ($m_d = 10^{-5}$ g to $10^{10} \, {\rm g}$), the aggregates are compressed mainly due to the
gas drag (\S 2.2.2), and therefore $\rho_{int}$ gradually increases\footnote{The effect of the collisional compression
is negligibly small in this case.}.
{After $S_t \gtrsim 1$, the aggregates behave as a $N$-body system 
under the effect of the turbulent fluctuation (eq. (\ref{eq:19}))}.
For $m_d > 10^{10}$ g and $S_t > 1$, the aggregates are compressed due to their self-gravity. 
 In the model shown in Fig. \ref{fig:2}, 
it becomes  gravitationally unstable at $t = 75$ Myr (see also \S 4.1).

Figure \ref{fig:2}b plots the collisional velocity $\Delta v$ of the aggregate and its size $a_d$ as a function of $m_d$. 
The size $a_d$ 
monotonically increases. In the compression stage ($m_d > 10^5$ g), the increase of $a_d$ slows down (see eq.(\ref{eq:10})).
Initially $\Delta v$ is 
20 cm s$^{-1}$, and it slightly decreases 
during the hit-and-stick stage. 
Then it turns to an increase phase  until $\sim 57$ m s$^{-1}$
around $S_t = 1$ during the compression stage. 
The size of the aggregate becomes $a_d \sim 10^4$ cm at the end of this stage.
In this case, the aggregates are not compressed by their self-gravity ($m_d < 10^{10}$ g) before the dust disk becomes GI.
Note that the increase of $\Delta v$ slows down at $m_d \sim 0.1$ g, which corresponds to 
the transition between $\Delta v_I$ and $\Delta v_{II} $ (eqs. (\ref{eq: deltav1}) and (\ref{eq: deltav2})).

\begin{figure}[h]
\begin{center}
\includegraphics[width = 18cm]{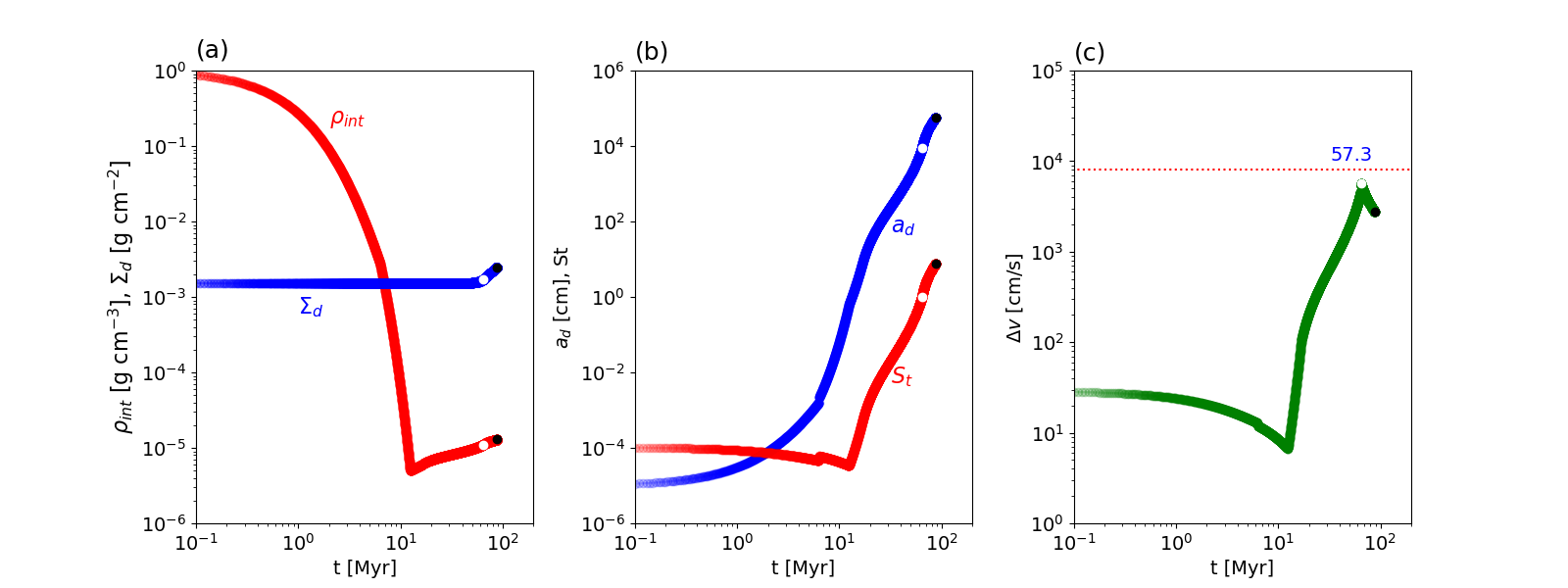} 
\caption{Time evolution of the internal density of the aggregate $\rho_{int}$, size $a_d$
and collision velocity $\Delta v$ for the same model shown in Fig. \ref{fig:2}.  
Evolution prior to the gravitational instability is plotted. 
The position where $S_t = 1$ for each quantity is shown by a white filled circle.
The maximum $\Delta v$ is shown (57.3 m s$^{-1}$ in this case) and
the critical velocity for the collisional destruction (i.e., 80 m s$^{-1}$) 
is shown by the red-dotted line in panel (c). 
 }
\label{fig:3}
\end{center}

\end{figure}

\begin{figure}[h]
\begin{center}
\includegraphics[width = 18cm]{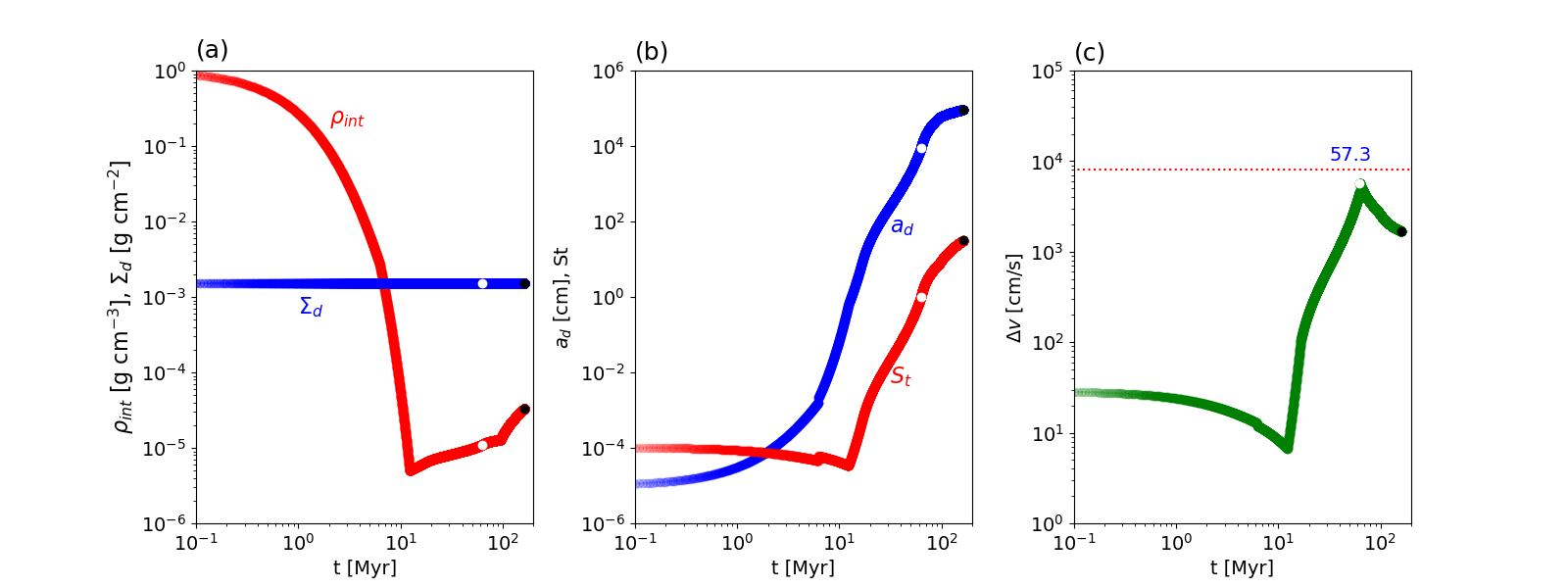} 
\caption{Same as Fig. \ref{fig:3}, but the model \textit{without} the {radial drift} of the dust.
Note that the time of GI is 136 Myr, in contrast to 71 Myr in Fig. \ref{fig:3}.
 }
\label{fig:noadvec}
\end{center}

\end{figure}

%
%

The growth time of an aggregate for $\Delta v = \Delta v_I \simeq  1/2 \sqrt{\alpha} c_s R_e^{1/4} S_{t} $ can be estimated as 
\begin{eqnarray}
t_{grow} &\equiv &\left(\frac{d \ln m_d}{dt} \right)^{-1}  \label{eq: 6} \nonumber  =  \frac{4\sqrt{2\pi}}{3}  \frac{H_d \, \rho_{int} \, a_d}{ \Delta v \, \Sigma_d} \\
&\simeq& \frac{16}{3} \sqrt{\frac{2}{\pi} }  \frac{H_g}{\sqrt{\alpha} \, R_e^{1/4} \, c_s \, f_{dg}}  \label{eq: 7} \nonumber \\
&\simeq&  2.9 \times 10^7 \; [{\rm yr}]  \;   c_{s,1}^{-1} \left( \frac{f_{dg}}{0.01} \right)^{-1} \left( \frac{H_g}{0.1 \, {\rm pc}}  \right)  \left( \frac{\alpha}{0.02}  \right)^{-1/2} \,   \left( \frac{R_e}{10^4}  \right)^{-1/4}. \label{eq: growthtime} 
%
\end{eqnarray} 
Figure \ref{fig:3} shows time evolution of $\rho_{int}$, $a_d$,  $S_t$, and $\Delta v$ for the same model shown
in Fig. \ref{fig:2}.  The hit-and-stick phase lasts for $\sim 10$ Myr as expected by $t_{grow}$,  and
$S_t$ becomes unity at $t = 60$ Myr.
At this moment, the dust aggregate's size reaches $\sim$ 100 m (Fig. \ref{fig:3}b). 

Fig.  \ref{fig:3}b shows that the growth of $a_d$ is exponential, or slower in time, 
in contrast to the results in Paper I. 
This is a natural consequence of the evolution of the mass of the dust aggregates.
The mass increase rate of the dust is 
\begin{eqnarray}
\frac{d m_d}{dt} \sim n_d \,a_d^2 \, \Delta v  \sim \frac{\Sigma_d}{H_d} \,a_d^2 \, \Delta v \propto S_t^{1/2} \, a_d^2 \, \Delta v.
\end{eqnarray}
Here we assume the dust layer is sedimanted, i.e., its thickness $H_d$ is scaled as $H_d \propto S_t^{-1/2}$.
For $\Delta v \propto S_t^{1/2}$ (eq. (\ref{eq: deltav2})),  $d m_d/dt \propto m_d$, 
therefore, $m_d$ grows exponentially.
The runaway growth seen in Paper I is caused by the assumption that 
 the internal density of the dust aggregates stays porous (i.e., the fractal dimension is $\sim 2$) through the evolution.
 In reality, when the compression by the ambient gas works, $\rho_{int}$ is nearly constant (i.e., $m_d \propto a_d^3$), as shown in Fig. \ref{fig:2}, therefore $S_t \propto a_d$.
If the scale height of the dust disk is constant, then $d m_d /dt \propto S_t^{1/2} a_d^2  \propto m_d^{5/6}$; therefore, the growth of 
the dust aggregate should be slower than $\exp(t)$.

Fig. \ref{fig:3}c shows that the collisional velocity $\Delta v$ gradually decreases during the hit-and-stick stage, 
and it turns to rapid increase during the compression stage until $S_t$ becomes unity at $t = 56$ Myr.
In the $N$-body stage, the collisional velocity decreases from its maximum value, 57 m s$^{-1}$, 
and it becomes GI (i.e., $Q_d \le 2$) at $t \simeq 75$ Myr.

For comparison, the evolution of the model without the {radial drift} is shown in Fig. \ref{fig:noadvec}. 
We found that the dust aggregates before $S_t = 1$ evolve almost the same way as that in the model with the {radial drift} (Fig. \ref{fig:3}).
However, the time for GI is 136 Myr, in contrast to 71 Myr for the case with the advection.

\begin{figure}[h]
\begin{center}
\includegraphics[width = 12cm]{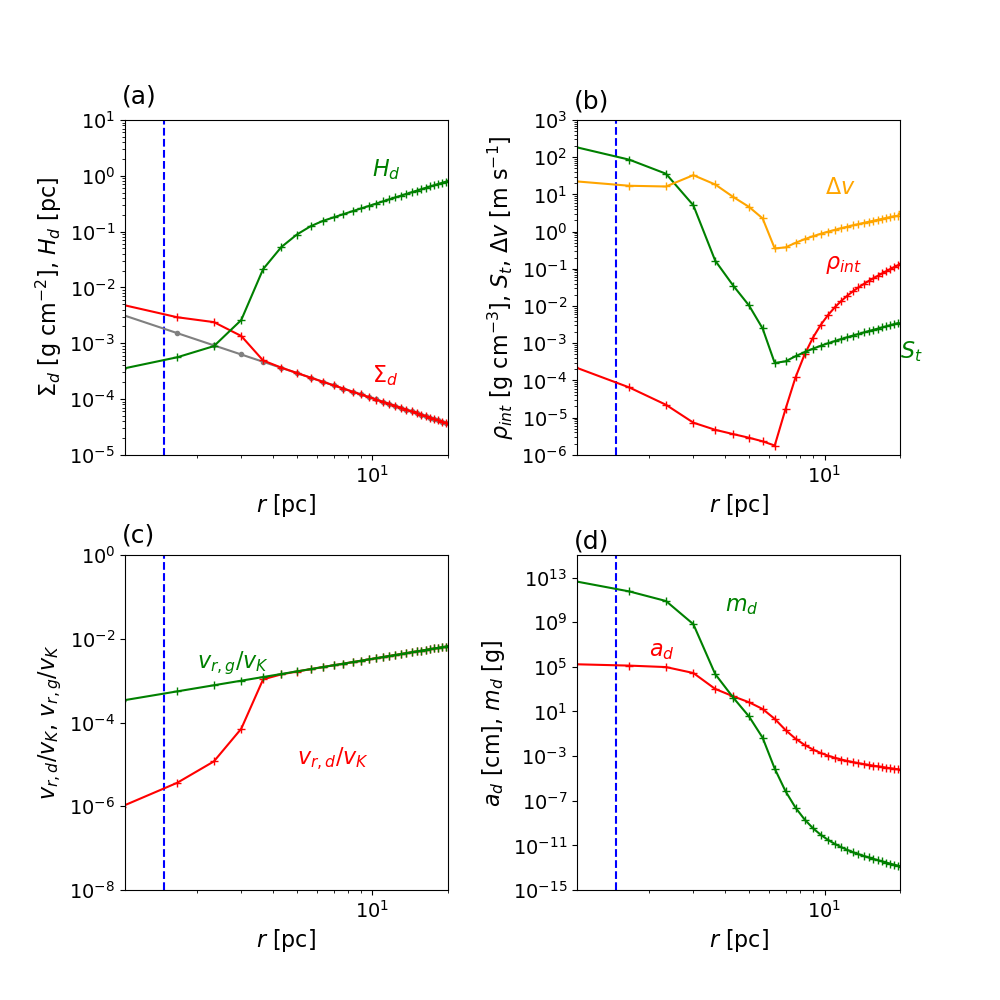} 
\caption{Radial distribution of various quantities at $t = 200$ Myr for the same model shown in Fig. \ref{fig:2}.  (a) Surface density of 
dust $\Sigma_d$ and the scale height of the dust $H_d$.  The gray line is $\Sigma_d$ at $t=0$. 
Note that the surface density of the dust decreases in the outer disk ($r  > 30$ pc), and the total 
mass of the dust is conserved.
The vertical dashed line is 
the position of the snowline $r_{snow} = 1.5 $ pc.
(b) Same as (a), but for the collision velocity $\Delta v$, 
the internal density of the aggregate $\rho_{int}$ and the Stokes parameter $S_t$.  (c) Same as (a), but for the radial velocity of the dust $v_{r, d}$ and 
$v_{r, g}$ normalized by the Kepler velocity $v_K$. (d). Same as (a), but for the mass and size of the aggregate, $m_d$ and $a_d$.  }

\label{fig:6}
\end{center}

\end{figure}

Figure \ref{fig:6} shows the radial distributions of $\Sigma_d, H_d, \Delta v$, the radial velocity of the dust and gas ($v_{r, d}$ and $v_{r, g}$), $\rho_{int}, S_t, m_d$, and $a_d$
at 200 Myr in the same model shown in Fig. \ref{fig:2}.
As Fig. \ref{fig:6}a shows, the dust is accumulated around $r \sim 2-3$ pc, 
where $v_{r,d} \ll v_{r,g}$ (Fig. \ref{fig:6}c) and 
$S_t$ turns from  $S_t < 1$ to $S_t > 1$ (Fig. \ref{fig:6}b).
From Fig. \ref{fig:6}d, the dust aggregates evolve more rapidly in the inner region ($r \lesssim 3$ pc), and the
maximum size is $\sim$ km. We call these objects as \textit{blanetesimals}.

%
\section{Discussion}
%

\subsection{Dependence on $\alpha$ and $M_{BH}$.}

\begin{figure}[h]
\begin{center}
\includegraphics[width = 14cm]{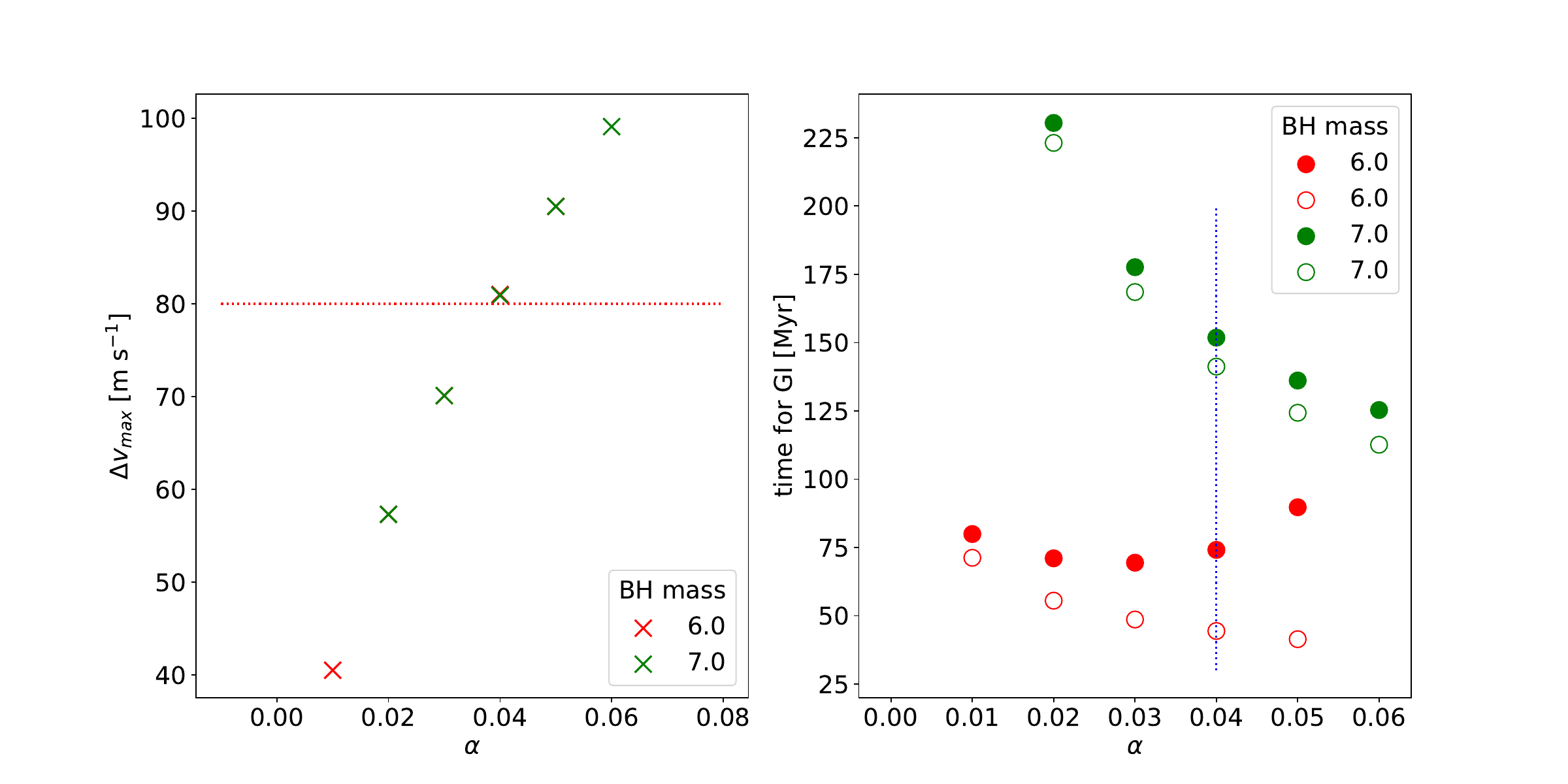} 

\caption{(a) $\Delta v_{max}$  as a function of $\alpha$ in the models with the {radial drift}. 
Red and  green crosses are $M_{BH} = 10^6 M_\odot$ and $10^7 M_\odot$, respectively.
The dotted line is the velocity limit for 
collisional destruction ($80$ m s$^{-1}$). (b)  Time for the gravitational instability (GI) as a function of $\alpha$. The filled circles are the time for GI and
the open circles are the time for $S_t = 1$.  
Red and  green circles are $M_{BH} = 10^6 M_\odot$ and $10^7 M_\odot$, respectively.
The $\alpha$ corresponds to $\Delta v_{max} = 80$ m s$^{-1}$, i.e. $\alpha = 0.04$ is shown by the blue dotted line. }
\label{fig:7}
\end{center}
\end{figure}

In the models with the {radial drift} of the dust aggregates, 
we investigated how the maximum velocity of the collision $\Delta v_{max}$ and the
time for GI ($\tau_{GI}$) depend on $\alpha$ and $M_{BH}$.
   In Figure \ref{fig:7}, we plot $\Delta v_{max}$ and $\tau_{GI}$ as 
a function of $\alpha$ for $M_{BH} = 10^6 M_\odot$ and $10^7 M_\odot$.
It shows that $\Delta v_{max}$ depends on $\alpha$, and not on $M_{BH}$. 
If $\Delta v_{max} \lesssim 80$ m s$^{-1}$ is necessary for collisional growth as numerical experiments suggested \citep{wada-k2009},
 then $\alpha$ should be $\sim 0.04$ or smaller.

The behavior of the dust growth (e.g., Fig. \ref{fig:2}) does not significantly depend on  $\alpha$ and $M_{BH}$, 
but the timescale to reach $S_t = 1$ is different as shown in Fig. \ref{fig:7}b.
 For example, for $M_{BH} = 10^6 M_\odot$ and $\alpha = 0.02$,
it takes $\sim 60$ Myr when $S_t$ exceeds unity, whereas it is $\sim 225$ Myr for $M_{BH} = 10^{7} M_\odot$ and $\alpha = 0.02$.
This implies that smaller BHs may preferentially host blanets within a lifetime AGNs ($ \lesssim10^8$ yr).
Figure \ref{fig:7}b shows that, for $M_{BH} = 10^7 M_\odot$,
the blanetesimal disk may not become GI  earlier than $\sim 150$ yr for $\alpha < 0.04$. 
For $\alpha > 0.05$ or 0.06, GI does not occur at $r = r_{snow}$ in the models with $M_{BH} = 10^6 M_\odot$ or $10^7 M_\odot$, respectively.

\subsection{Number and mass of \textit{blanets}}

\begin{figure}[h]
\begin{center}

\includegraphics[width =9cm]{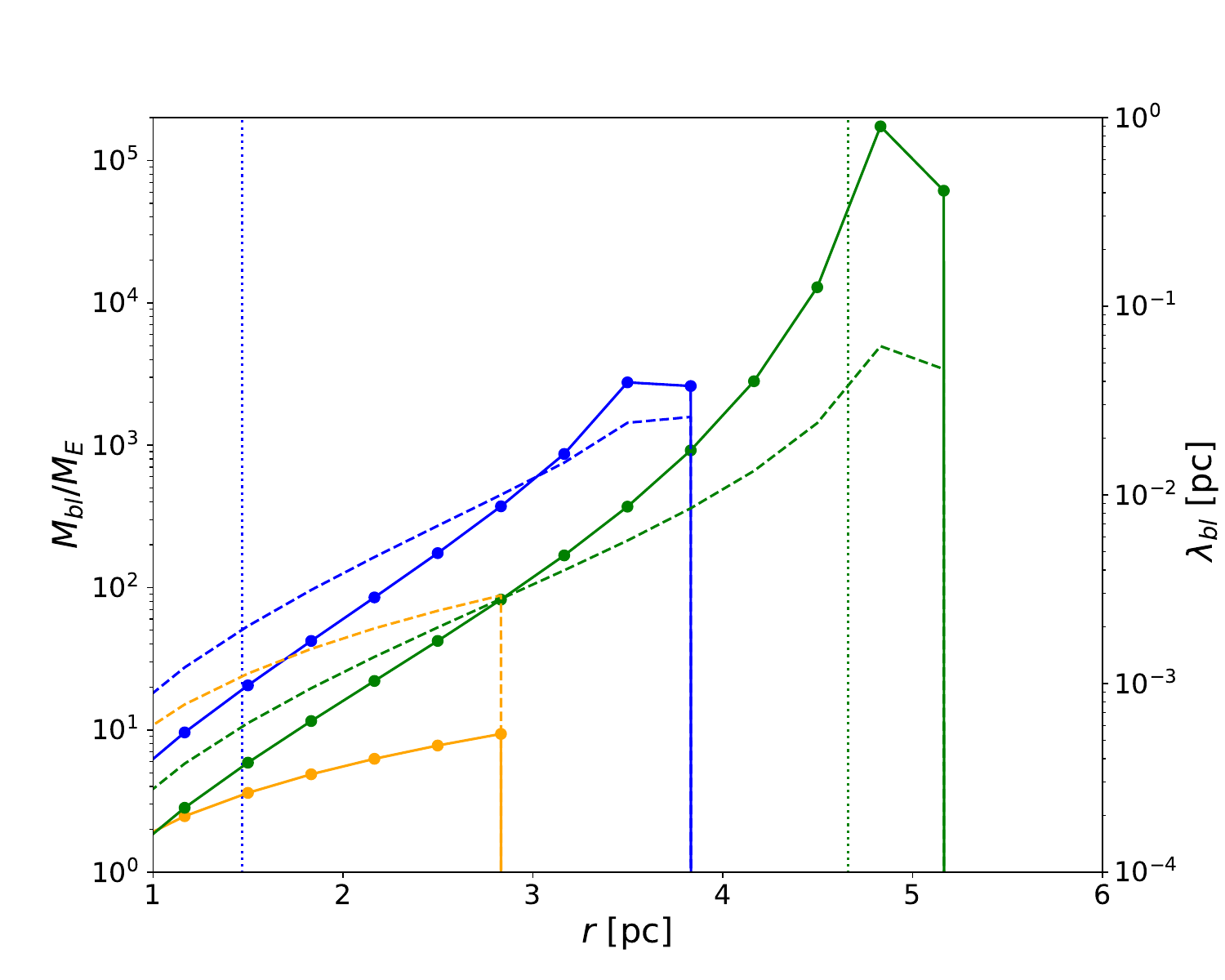} 

\caption{Radial distribution of mass in the Earth mass $M_E$ (solid lines) 
for the left vertical axis, and the critical wave length for GI ($\lambda_{bl}$) by the dashed lines for
the right vertical axis.
The advection models for $M_{BH} = 10^6 M_\odot, \alpha = 0.02$ (blue) and $M_{BH} = 10^7 M_\odot, \alpha=0.04$ (green) are shown
in blue and green lines, respectively.  
The model without advection ($\alpha = 0.02$) is shown by orange lines. The blue and green vertical dotted lines are 
position of snowlines for $M_{BH} = 10^6 M_\odot$ and  $M_{BH} = 10^7 M_\odot$, respectively.}
\label{fig:8}
\end{center}
\end{figure}

In the final stage of the evolution, the \textit{blanetesimal} disk can be gravitationally unstable,
 and it fragments into massive objects, i.e.,
\textit{blanets} (see \S 2.3).
Figure  \ref{fig:8}  shows the radial distribution of the mass and typical separation between blanets,  $\lambda_{bl} \approx \lambda_{GI}$ (eq.(\ref{eq: lambda_gi})). 
Two models with the {radial drift} for $M_{BH} = 10^6 M_\odot$ with $\alpha = 0.02$ and  $M_{BH} = 10^7 M_\odot$ with 
$\alpha = 0.04$ are shown.
For comparison, a model without the {radial drift} is also shown ($M_{BH} = 10^6 M_\odot$ and $\alpha = 0.02$).
The mass of blanets ranges from $\simeq 20M_E$ at $r = r_{snow}$ to $\simeq 3000 M_E$ at $r \sim 3.5$ pc 
for $M_{BH} = 10^6 M_\odot$, in contrast to the model  without advection, which is $M_{bl} \simeq 3 M_E -7 M_E$.
For  $M_{BH}  =10^7 M_\odot$, $M_{bl} \gtrsim 10^{4} M_E -10^5 M_E$ outside the snowline.
However, this extraordinary massive blanet is unlikely, because it is comparable to the minimum mass of brown dwarfs ($\sim 2\times 10^4 M_E$).
Therefore, the largest size of the blanets ($R_{bl}$) would be 
maximally $\sim  10 \times$ Earth's radius at $r \sim 3$ pc for $M_{BH} = 10^6 M_\odot$, if the average internal density is similar to that of the Earth. 

Number of blanets is $\sim 8\times 10^5$ at $r = r_{snow}$ for $M_{BH} = 10^6 M_\odot$ with $\alpha = 0.02$,
{provided that all the dust is converted to blanets through GI.  }
From  Fig. \ref{fig:8}, the average distance between blanets would be 
$\lambda_{bl} \sim 10^{-3}$ pc at $r = r_{snow}$.
Therefore, the system of blanets does not resemble any known exoplanet systems, in a sense that 
the ``planets'' are isolated, dominant objects in their orbits.

{Is the swarm of blanets  the final form of the system?   The collisional time scale $t_{coll}$ for the blanets can 
be estimated as:
\begin{eqnarray}
t_{coll}^{-1} =  4 \sqrt{\pi} \, n_{bl} \, \sigma_{bl} \left( r_{coll}^2 + \frac{G \,m_{bl}}{\sigma_{bl}^2} r_{coll}  \right), 
\label{eq:col}
\end{eqnarray}
where where $n_{bl}$ and $\sigma_{bl}$ are the number density and velocity dispersion
of the blanets, respectively.  $r_{coll}$ is the distance at the closest approach, which is $r_{coll} \sim 2 R_{bl}$   \citep{binney2008}.
The second term of eq. (\ref{eq:col}) represents  the gravitational focusing, which enhances the
collision rate.
For  $n_{bl} \sim \lambda_{bl}^{-3} \simeq (10^{-3} \, {\rm pc} )^{-3}$, $M_{bl} \simeq 20 M_E$, $R_{bl} \simeq 3 R_E$,  and $\sigma_{bl} \sim \Delta v \simeq 30$ m s$^{-1}$ (Fig. \ref{fig:3}c),
the collisional time is $t_{coll} \approx 6$ Gyr.  This can be shorter, if three-body encounters
between a close-binary and a blanet are considered.  
Therefore, the blanet system could dynamically evolve in the cosmological time scale, and 
it would be interesting study the evolution by direct $N$-body simulations.}


\subsection{Can blanets acquire a massive gas envelope? }
{
It would be interesting to investigate whether blanets can obtain an atmosphere.  
If the Hill radius $r_H$ of a blanet is larger than the disk scale height, the blanets may form a gap, 
then the obtained mass of the atmosphere depends on $r_H$.
The Hill radius of a blanet,  $r_H = (M_{bl}/3 M_{BH})^{1/3} \, r$  is
\begin{eqnarray}
r_H \simeq  10^{-4} \, {\rm pc} \left(\frac{M_{bl}}{1000 \,M_E}  \right) ^{1/3}, 
\end{eqnarray}
whereas  the scale height of the gas disk at $r = 1.5 $ pc for $M_{BH} = 10^6 M_\odot$ is,  $H_g \simeq c_s/\Omega_K \simeq 2 \times 10^{-2} M_{BH,6}^{-1/2} $ pc,
which is larger than $r_H$.}

{On the other hand,  the Bondi radius is
\begin{eqnarray}
r_B \equiv \frac{GM_{bl}}{c_s^2}  \simeq  1.3 \times 10^{-5} \, {\rm pc} \left( \frac{M_{bl}}{1000\, M_E}  \right)  \left(\frac{c_s}{1 \, {\rm km}  \, {\rm s}^{-1} } \right)^{-2}.
\end{eqnarray}
Therefore, we suspect that the mass of the atmosphere is limited by the Bondi accretion, rather than by a gap formation.
If this is the case, the envelope mass would be 
%
 \begin{eqnarray}
 M_{env} &\sim&  \frac{4\pi}{3} r_B^3 \, \rho_g \\
 & \approx & 10^{-7} M_E  \left( \frac{M_{bl}}{1000\, M_E}  \right)^3  \left(\frac{c_s}{1 \, {\rm km} \, {\rm s}^{-1}} \right)^{-6} \left(  \frac{\rho_g}{2\times 10^{-21} \, {\rm g} \, {\rm cm}^{-3} }  \right).
 \end{eqnarray}
However,  if the Bondi radius is filled with the ambient gas during the orbital motion of a blanet,
 the envelope mass could be maximally 
 \begin{eqnarray}
 M_{env} &\sim&  2 \pi \, r_{bl} \cdot \, \pi r_B^2  \, \rho_g  \sim 0.05 M_E,
 \end{eqnarray}
 for $M_{bl} = 1000 M_E$ at $r_{bl} = 1.5 $ pc.
The small $M_{env}$ would suggest that the runway accretion of the gas to blanets 
and formation of massive ``gas giants''  are difficult, but it also depends on how quickly are the orbits of blanets filled with the gas.
This is an interesting open question for future studies.
}
 

\subsection{Can other mechanisms of planet formation be applicable?}
{
In this paper, we focused on the evolution of the dust aggregates 
based on the coagulation theories of dust monomers and the gravitational instability. 
Other formation mechanisms of planetesimals have also been proposed and 
extensively discussed in the planet formation community.
Among them,  we here look over the pebble accretion, secular gravitational instability (secular GI),
and the streaming instability as possible mechanisms of blanetesimal formation. 
More quantitative analysis in the circumnuclear environment
would be interesting for future studies.}

{
The pebble accretion
is an accretion process of small solid bodies (i.e., pebbles) 
to massive seed objects (e.g., planets or planetesimals) under the effect of
the gas drag and gravity \citep[e.g.,][]{ormel2017,  lambrechts2019}.
In the present case, relatively massive aggregates could accumulate ambient smaller particles, then they could become more
massive. However,  the time scale conditions for the pebble accretion \citep{ormel2018}, i.e., 
$ t_{settl} < t_{enc}$ and $t_{stop}  < t_{enc}$ are not likely to be satisfied in the 
circumnulcear disk, 
where the settling time $t_{settl}$ is the time needed for a particle to sediment to the massive objects, 
and  the encounter time $t_{enc}$ is the duration of the gravitational encounter time.
In other words, the radial flux of pebbles due to the gas drag is too small in the region of $S_t \ll 1$. 
Moreover, the relatively large turbulent motion of the gas (i.e., $\alpha \gtrsim 0.01$) may prevent from 
the pebble accretion \citep{ormel2018}, in contrast to the circumstellar disk.
Therefore, we do not expect that the pebble accretion is a major process as a formation mechanism of 
blanets.
}

{
The secular GI, which is the gravitational instability due to gas-dust friction, 
is another possible mechanism to form 
planets in the circumstellar disk \citep{youdin2011}. 
According to \citet{takahashi2014}, 
the condition for this instability is expressed as
\begin{eqnarray}
\Gamma \equiv
\left(\frac{\alpha}{4\times 10^{-5}} \right)
\left( \frac{f_{dg}}{0.1} \right)^{-2}
\left(\frac{Q_g}{10}  \right)
\left(\frac{\eta}{0.01}  \right)  \lesssim 1.
\end{eqnarray} 
In the present case, $\Gamma \sim 100$ for $\alpha \gtrsim 0.01$, $f_{dg} = 0.01$,  $\eta \sim 10^{-4}$ and $Q_g = 2$, 
therefore we do not expect the secular GI in the circumnuclear disk. 
}

{The streaming instability could be an effective mechanism to make condensations of 
dust particles,  if the dust-to-gas mass ratio ($f_{dg}$) is close to unity \citep[e.g.][]{youdin2005}.
According to numerical simulations by \citet{carrera2015}, coagulation of particles driven by the streaming instability
depends on $f_{dg}$ and $S_t$. They found that the streaming instability occurs for $f_{dg} \sim 0.02$ for $S_t \sim 0.1$.
In our case, the dust-to-gas mass ratio increases to  $f_{dg} \simeq 0.02-0.03$ from the initial value (0.01) outside the snowline due to the radial drift of the dust aggregates. However,  $S_t \gg 1$ in the region (Fig. \ref{fig:6}b), therefore we do not expect that the streaming instability occurs in the circumnuclear disk.
}

\section{Summary}
In this follow-up paper of \citet{wada2019} (Paper I), we theoretically investigated a process of dust evolution around a SMBH in the galactic center.
{We proposed that a new class of astronomical objects, \textit{blanets}
can be formed, provided 
that the standard scenario of planet formation is present in the circumnuclear disk.
}

Here, we investigated the physical conditions of the blanet formation outside the snowline ($r_{snow} \sim $ several parsecs) in more detail, especially considering the effect of the {radial drift} of the dust aggregates. 
We also improved the dust evolution model in Paper I in terms of the internal density evolution of the dust aggregates.
We assumed the maximum collisional velocity for destruction, which was suggested by previous numerical simulations, as one of necessary conditions for
blanet formation. 
{We found that a dimensionless parameter $\alpha = v_t^2/c_s^2$, where $v_t$ is the turbulent velocity and $c_s$ is the sound velocity, describing the effective angular momentum transfer due to the turbulent viscosity in the circumnuclear disk 
should be smaller than 0.04}
 for the black hole mass $M_{BH} = 10^{6} M_\odot$; otherwise, the dust aggregates could be destroyed due to collisions.
The formation timescale of blanets $\tau_{GI}$ at $r_{snow}$ is $\tau_{GI} \simeq$ 70-80 Myr for $\alpha = 0.01-0.04$.
The blanets ($M_{bl}$) are more massive for larger radii; they range from $M_{bl} \sim 20 M_E - 3000 M_E$ in $r < 4$ pc,
in contrast to  $M_{bl} = 3-7 M_E$ for the case without the {radial drift}.  

The typical separations between the blanets, estimated from the wavelength of the gravitational instability, would be $\sim 10^{-3} - 10^{-2}$ pc.  

For $M_{BH} \ge 10^7 M_\odot$,  the formation timescale is longer than $\sim 150$ Myr for $\alpha \le 0.04$. 
Although the gravitational instability of the blanetesimal disk takes place just outside the snowline ($r = 4.7$ pc), they should not be blanets
because  they are more massive than the
typical brown dwarf mass ($\sim 3\times 10^4 M_E$). 
Note that AGNs are often heavily obscured with dense gas even for hard X-rays \citep{buchner2014} ($N_{\rm H} > 10^{23}$ cm$^{-2} $, i.e., Compton-thick).
If this is the case, the snowline is located at the inner region (e.g. $r \sim 2-3 $ pc), and as a result, blanets with $M_{bl} \sim 10 M_E-100 M_E$ 
around $M_{BH} = 10^7 M_\odot$ could be possible.
Our results suggest that  \textit{blanets} could be formed around relatively low-luminosity AGNs during their lifetime ($\lesssim 10^8$ yr). The system of \textit{blanets} should be extraordinarily different from the standard Earth-type planets in the exoplanet systems. {The blanets may acquire a rarefied atmosphere due to accretion of the gas in the circumnuclear disk (\S 4.3).}
{The dynamical evolution of the swarm of blanets around a SMBH and 
whether they become more massive objects or destroyed due to collisions may be an interesting subject for future studies (\S 4.2).
}

\acknowledgments
We would like to appreciate the anonymous referee's many valuable comments.
The authors also thank Hidekazu Tanaka for many thoughtful comments.
This work was supported by JSPS KAKENHI Grant Number  18K18774.  






\newpage

\begin{thebibliography}{}
\bibitem[Barvainis(1987)]{barvainis1987} Barvainis, R.\ 1987, \apj, 320, 537
\bibitem[Buchner et al.(2014)]{buchner2014} Buchner, J. Georgakakis, A., Nandra, K. et al.   A\&Ap 564, 125 (2014)
\bibitem[Binney, \&Tremaine(2008)]{binney2008} Binney, J., Tremaine, S. "Galactic Dynamics 2nd. Edition", p. 626, Princeton University Press, Princeton (2008)
\bibitem[Carrera et al.(2015)]{carrera2015} Carrera, D., Johansen, A., \& Davies, M.~B.\ 2015, \aap, 579, A43. doi:10.1051/0004-6361/201425120
\bibitem[Greenhill et al.(2003)]{greenhill2003} Greenhill, L.~J., Booth, R.~S., Ellingsen, S.~P., et al.\ 2003, \apj, 590, 162. doi:10.1086/374862
\bibitem[Izumi et al.(2018)]{izumi2018} Izumi, T., Wada, K., Fukushige, R., et al.\ 2018, \apj, 867, 48
\bibitem[Kataoka et al.(2013)]{kataoka2013} Kataoka, A., Tanaka, H., Okuzumi, S., et al.\ 2013, \aap, 557, L4
\bibitem[Kudoh et al.(2020)]{kudoh2020} Kudoh, Y., Wada, K., \& Norman, C.\ 2020, arXiv:2008.07050
\bibitem[Lambrechts et al.(2019)]{lambrechts2019} Lambrechts, M., Morbidelli, A., Jacobson, S.~A., et al.\ 2019, \aap, 627, A83. doi:10.1051/0004-6361/201834229
\bibitem[Michikoshi, \& Kokubo(2017)]{michikoshi2017} Michikoshi, S., \& Kokubo, E.\ 2017, \apj, 842, 61
\bibitem[Michikoshi, \& Kokubo(2016)]{michikoshi2016} Michikoshi, S., \& Kokubo, E.\ 2016, \apjl, 825, L28
\bibitem[Mukai et al.(1992)]{mukai1992} Mukai, T., Ishimoto, H., Kozasa, T., et al.\ 1992, \aap, 262, 315
\bibitem[Netzer(2015)]{netzer2015} Netzer, H.\ 2015, \araa, 53, 365
\bibitem[Okuzumi et al.(2009)]{okuzumi2009} Okuzumi, S., Tanaka, H., \& Sakagami, M.-. aki .\ 2009, \apj, 707, 1247
\bibitem[Okuzumi et al.(2012)]{okuzumi2012} Okuzumi, S., Tanaka, H., Kobayashi, H., et al.\ 2012, \apj, 752, 106
\bibitem[Ormel(2017)]{ormel2017} Ormel, C.~W.\ 2017, Formation, Evolution, and Dynamics of Young Solar Systems, 197. 
\bibitem[Ormel, \& Cuzzi(2007)]{ormel2007} Ormel, C.~W., \& Cuzzi, J.~N.\ 2007, \aap, 466, 413
\bibitem[Ormel \& Liu(2018)]{ormel2018} Ormel, C.~W. \& Liu, B.\ 2018, \aap, 615, A178. doi:10.1051/0004-6361/201732562
\bibitem[Sato et al.(2016)]{sato2016} Sato, T., Okuzumi, S., \& Ida, S.\ 2016, \aap, 589, A15
\bibitem[Schartmann et al.(2014)]{schartmann2014} Schartmann, M., Wada, Keiichi, Prieto, M.~A., Burkert, A., \& Tristram, K.~R.~W.\ 2014, \mnras, 445, 3878 
\bibitem[Shlosman, \& Begelman(1987)]{shlosman1987} Shlosman, I., \& Begelman, M.~C.\ 1987, \nat, 329, 810
\bibitem[Suyama et al.(2008)]{suyama2008} Suyama, T., Wada, K., \& Tanaka, H.\ 2008, \apj, 684, 1310
\bibitem[Suyama et al.(2012)]{suyama2012} Suyama, T., Wada, Koji, Tanaka, H., et al.\ 2012, \apj, 753, 115
\bibitem[Takahashi \& Inutsuka(2014)]{takahashi2014} Takahashi, S.~Z. \& Inutsuka, S.-. ichiro .\ 2014, \apj, 794, 55. doi:10.1088/0004-637X/794/1/55
\bibitem[Tsukamoto et al.(2017)]{tsukamoto2017} Tsukamoto, Y., Okuzumi, S., \& Kataoka, A.\ 2017, \apj, 838, 151
\bibitem[Wada et al.(2008)]{wada-k2008a} Wada, Koji., Tanaka, H., Suyama, T., et al.\ 2008, \apj, 677, 1296
\bibitem[Wada \& Norman(2002)]{wada2002b} Wada, K. \& Norman, C.~A.\ 2002, \apjl, 566, L21. doi:10.1086/339438

\bibitem[Wada(2012)]{wada2012} Wada, Keiichi. 2012, \apj, 758, 66 
\bibitem[Wada(2015)]{wada2015} Wada, Keiichi.\ 2015, \apj, 812, 82 
\bibitem[Wada et al.(2009)]{wada-k2009} Wada, Koji., Tanaka, H., Suyama, T., et al.\ 2009, \apj, 702, 1490
\bibitem[Wada et al.(2013)]{wada-k2013} Wada, K., Tanaka, H., Okuzumi, S., et al.\ 2013, \aap, 559, A62
\bibitem[Wada et al.(2016)]{wada2016} Wada, Keiichi., Schartmann, M., \& Meijerink, R.\ 2016, \apjl, 828, L19 
\bibitem[Wada et al.(2018)]{wada2018} Wada, Keiichi., Fukushige, R., Izumi, T., \& Tomisaka, K.\ 2018, \apj, 852, 88 
\bibitem[Wada et al.(2019)]{wada2019} Wada, K., Tsukamoto, Y., \& Kokubo, E.\ 2019, \apj, 886, 107 (Paper I)
\bibitem[Weidenschilling(1977)]{weidenschilling1977} Weidenschilling, S.~J.\ 1977, \mnras, 180, 57
\bibitem[Youdin \& Goodman(2005)]{youdin2005} Youdin, A.~N. \& Goodman, J.\ 2005, \apj, 620, 459. doi:10.1086/426895
\bibitem[Youdin, \& Lithwick(2007)]{youdin2007} Youdin, A.~N., \& Lithwick, Y.\ 2007, \icarus, 192, 588
\bibitem[Youdin(2011)]{youdin2011} Youdin, A.~N.\ 2011, \apj, 731, 99. doi:10.1088/0004-637X/731/2/99

%

\end{thebibliography}
\end{document}